# A Multilayer Comparative Study of XG-PON and 10G-EPON Standards

Charalampos Konstadinidis[1], Panagiotis Sarigiannidis[2], Periklis Chatzimisios[1], Paschalis Raptis[1], Thomas D. Lagkas[3]

[1]Department of Informatics, Alexander TEI, Thessaloniki, Greece
e-mail: {chrkon, peris, praptis}@it.teithe.gr
[2]Department of Informatics and Telecommunications Engineering, University of Western Macedonia, Kozani, Greece
e-mail: psarigiannidis@uowm.gr
[3]Computer Science Department, The University of Sheffield Internationally Faculty, CITY College, Thessaloniki, Greece
e-mail: tlagkas@city.academic.gr

**Abstract.** The purpose of this paper is to provide a multilayer review of the two major standards in next generation Passive Optical Networks (PONs) and technologies, the ITU-T 10-Gigabit-capable PON (XG-PON) and the IEEE 10 Gigabit Ethernet PON (10G-EPON). A study and a discussion on the standards are performed. The main intention of this paper is to compare XG-PON and 10G-EPON, mainly in terms of physical and data link layers**.** The paper answers the question of what are the common elements and the basic execution differences of the two standards. Moreover, critical points are raised regarding the Dynamic Bandwidth Allocation (DBA) schemes of both standards. Special focus is also pointed in the coexistence of XG-PON and 10G-EPON. Finally, the paper includes a discussion of open issues and continuing research regarding the two standards.

## 1 Introduction

Optical networks penetrate access networks in an increasing rate attributed by the extensive growth of Internet traffic as well as the increased use of bandwidth intensive applications. Cutting-edge multimedia services, such as Ultra High Definition (HD) video, are leading to even higher bandwidth requirements. Compared to copper, optical fiber can provide higher bandwidth over a longer distance. In order to partially fulfill the aforementioned requirement, telecommunication companies have been already deploying Fiber-To-The-x (FTTx) networks in various parts of the world. The concept of the different variations of the FTTx technology, namely FTTH (Home), FTTC (Curb) and FTTB (Building) is the provision of fast fiber connections in high proximity to the end user's premises. FTTx networks primarily constitute Passive Optical Networks (PONs) that are considered as one of the most promising alternatives to dominate broadband access, due to their cost-effective and practically limitless bandwidth potentials. PONs are widely deployed in FTTx technologies, creating optical light-paths without incorporating optical-to-electrical conversions [1]. A generic view of a PON is depicted in Fig. 1.

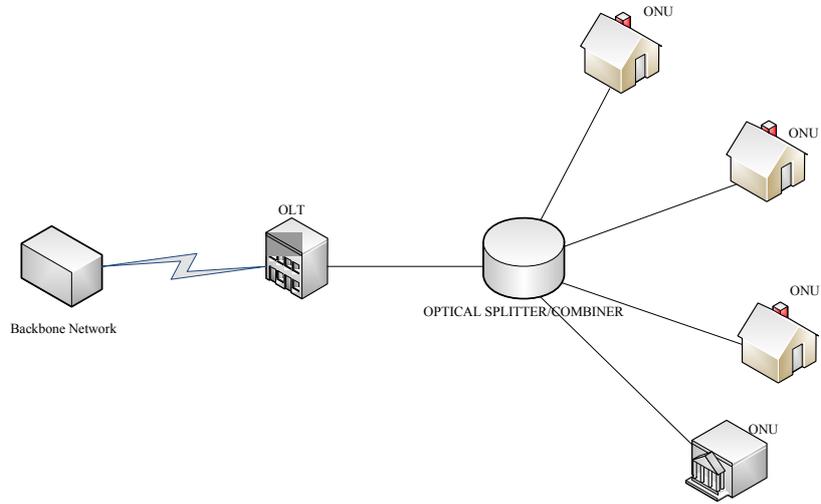

**Fig. 1.** The generic tree topology of a PON.

ITU-T Gigabit-capable PON (GPON) and IEEE PON (EPON) are the two competing systems since 2009 having both of them the advantage to offer more bandwidth per subscriber than their predecessor Broadband PON (BPON). EPON has been proved a success mainly in countries in Eastern Asia, such as China, Korea and Japan, while GPON has been widely deployed in the North America.

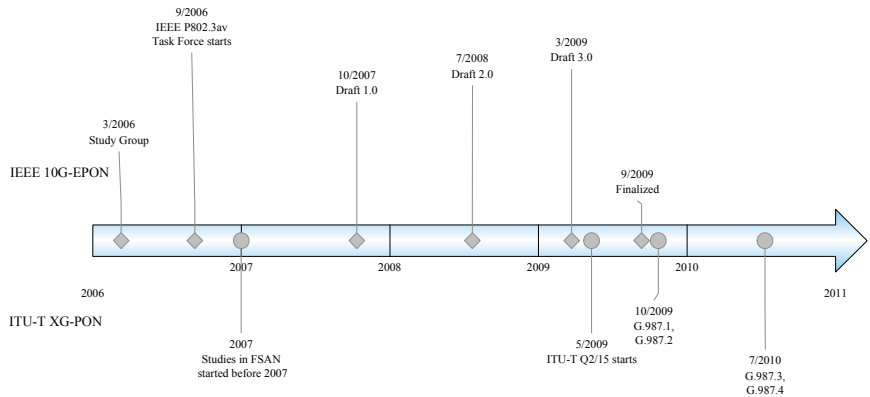

**Fig. 2.** Standardization timeline for 10 Gigabit PONs in IEEE and ITU-T [3]

### 1.1 10 Gigabit PONs

PON systems of 10 Gbit/s have been recently introduced to cover future high bandwidth demands and well as residential and backbone services [2,3]. The latest related IEEE and ITU-T standards are backwards compatible with previous PON generations, allowing upgrading in a progressive manner with minimal financial investment on the Optical Distribution Network (ODN) and minimal effect on existing users' operation. The standardization of IEEE 10G-EPON began in 2006 and ended in late 2009. On the other hand, ITU-T XG-PON has received its final form in 2010. The timeline for the standardization of IEEE 10G-EPON and ITU-T XG-PON is shown in Fig. 2.

IEEE and ITU-T along with the Full Services Access Network (FSAN) group have recently defined their 10 Gbit/s solutions, namely IEEE 802.3av, 10GE-PON, and ITU-T XG-PON, respectively, in an effort to support the upcoming bandwidth increase over existing ODNs. The P802.3av Task Force has been standardizing 10 G-EPON, developing the IEEE 802.3av 10GE-PON which provides symmetric (10 Gbit/s downlink and uplink) and asymmetric (10 Gbit/s downlink and 1 Gbit/s uplink) connections. The asymmetric traffic generated by IP video services is expected to be supported by the latter operation mode.

Regarding the related work of the FSAN group, its efforts are focused on examining next generation approaches that facilitate high provision, increased split ratio, and extensive reach. Under the consideration of Next Generation Access (NGA), which is mainly about technologies compatible with existing GPONs, XG-PON1 has been developed by ITU-T and FSAN; the supported asymmetrical downlink/uplink bandwidth capacity is 10/2.5 Gbit/s. The definition of XG-PON1 can be found in the ITU-T G.987.x series of recommendations. Some key characteristics of the XG-PON1 networks include security enhancements through authentication of management messages and power conservation through switching down parts or all of the Optical Network Units (ONUs) [4].

## 2 XG-PON Overview

After standardizing PON networks operating at 1 Gbit/s in ITU-T G.984.x Recommendation series, efforts have been made to standardize the 10 Gbit/s capable PONs (were finalized in October 2009 and published by ITU-T in March 2010). The ITU-T G.987.x series addresses the general requirements of 10 Gigabit capable passive optical networks in a way that promotes backward compatibility with the existing ODN that complies with the GPON systems. Furthermore, the XG-PON system utilizes Wavelength Division Multiplexing (WDM) defined in ITU-T G.984.x series and provides a seamless migration from Gigabit PON to XG-PON.

### 2.1 Physical Layer

The physical layer, also referred to as the Physical Media Dependent (PMD) layer is described in the ITU-T G.987.2 Recommendation and specifies a flexible optical fiber access network capable of supporting high bandwidth requirements. A general requirement of XG-PON is to provide higher data rates than GPON combined with minimized costs. Therefore, backward compatibility with legacy GPON deployments was a major topic in the physical layer specifications. To achieve the backward compatibility and co-existence of GPON and XG-PON systems, the optical wavelengths that selected for XG-PONs were the "O-band" (for the upstream ranging from 1260 to 1280 nm) and the "1577nm" (for the downstream ranging from 1575 to 1580 nm).

XG-PON systems are divided into XG-PON1 (featuring a 2.5 Gbit/s upstream path) and XG-PON2 (featuring a 10 Gbit/s one). ITU-T G.987.x Recommendation series [5,6,7] only addresses XG-PON1. It seems that a natural progression from GPON to XG-PON1 and XG-PON2 is to come when the technology becomes more mature.

Having in mind that XG-PON systems should share the same optical distribution network as GPONs, the ODN characteristics are quite clear (28 dB maximum loss in the windows from 1260 to 1360 and from 1480 to 1580 nm). The FSAN operators identified two loss budgets for the XG-PON systems. The Nominal 1 class, with a 29 dB maximum loss and the Nominal 2 class with a 31 db maximum loss. The first one allows XG-PON to coexist with standardized GPON systems, while the second one supports coexistence with the super-standard 29.5 dB GPON systems.

**2.2 Transmission Convergence (TC) Layer**

The XG-PON data link layer (usually referred as Transmission Convergence - TC layer) is composed of three distinct sublayers: the XGTC framing sublayer, the XGTC PHY adaption sublayer and the XGTC service adaption sublayer [8]. The TC layer's main function is to provide transport multiplexing between the Optical Line Terminal (OLT) and the ONU, followed by other functions like adaption of client layer signal protocols, Physical Layer Operations And Maintenance (PLOAM), interface for Dynamic Bandwidth Allocation (DBA), and ONU ranging and registration. XGTC functions are realized through Transmission CONTainers (T-CONTs) [9], each of them identified by a unique Allocation ID (Alloc-ID) assigned by the OLT.

The XGTC service adaption sublayer is responsible for taking the user's payload and formatting them for transmission over optical network. In an XG-PON system, the Service Data Units (SDUs), which include the user data frames and high-level PON management frames, are transmitted in XGTC payload section using the XG-PON Encapsulation Method (XGEM system). The XGEM supports SDU fragmentation, encapsulation and delineation both in the downstream and upstream directions, and marks the individual flows of traffic (ports) so that they can be accepted by the appropriate client on the other side of the PON.

The XGTC framing sublayer is responsible for the construction and parsing of the overhead fields in both the transmitter and the receiver side. On the transmitter side, the framing sublayer accepts the XGEM frames from the XGTC service adaption sublayer and then constructs the downstream XGTC frame or the upstream XGTC burst by providing embedded OAM and PLOAM messaging channel overhead fields. On the receiver side, the framing sublayer accepts the XGTC frames or XGTC bursts, parses the overhead fields by extracting the incoming OAM information and PLOAM

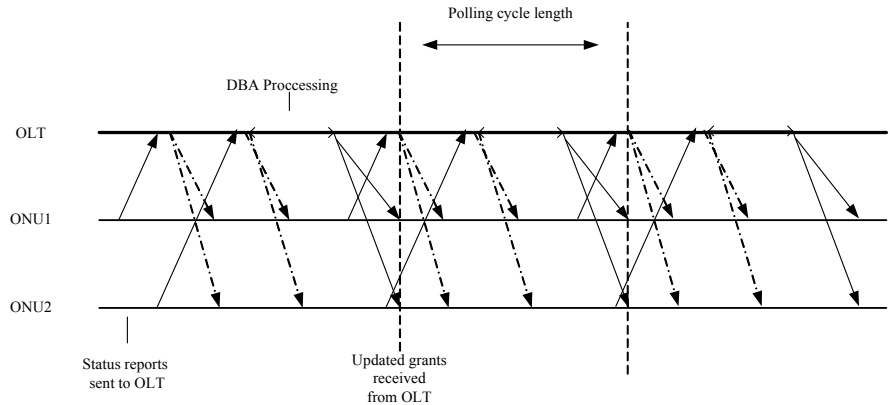

**Fig. 3.** The DBA process in XG-PON Systems.

messaging flows and then delivers the XGTC payloads to the service adaption sublayer.

The PHY adaption sublayer takes care of the low level coding of the TC frame over the physical channel trying to improve the detection, reception and delineation properties of the signal transmitted over the medium. Much of the work of the PHY adaption layer concerns the use of Forward Error Correction (FEC), a required feature for downstream and upstream directions. The use of FEC improves the effective sensitivity and overload characteristics of the optical receiver by introducing redundancy in the transmitted bit stream and allowing the receiver to operate under higher BER (Bit Error Rate) scenarios.

**XGTC Encapsulation Method**

The transmission of the SDUs (that include the user data frames and high-level PON management frames in the XTC payload section of the downstream XGTC frames and upstream XGTC bursts) is accomplished through the XGEM encapsulation method. The XGTC payload section is carried in the downstream frames or upstream bursts and contains one or more XGEM frames. Each XGEM frame contains a fixed size header carrying information such as Payload Length Indication (PLI), XGEM port-ID as well as last fragment indication and variable size XGEM payload field controlled by the PLI.

**Dynamic Bandwidth Allocation in XG-PON**

In a XG-PON system the OLT provides media access control for the upstream traffic. The basic idea is that each downstream frame comes with a BandWidth map (BWmap) that indicates the location for upstream transmissions by various ONUs in the corresponding upstream PHY frame. The header of downstream XGTC frame contains the BWmap field, which specifies a bandwidth allocation to a particular allocation ID (Alloc-ID), while in the upstream burst the allocation overhead is composed of the Dynamic Bandwidth Report upstream (DBRu) structure and contains the Buffer Occupancy (BuffOcc) field that reports the total amount of SDU traffic.

XG-PON uses Point-to-Multi-Point (P2MP) connections between the OLT and the ONU. Fig. 3 illustrates the logical diagram of the DBA process. Due to the high available bandwidth in XG-PON, bandwidth allocation is based on the Service Level Agreements (SLA) where Quality of Service (QoS) can be granted according to the demand. Bandwidth is allocated per Transmission CONTainer (T-CONT) [9,10], which is the basic control unit for bandwidth allocation. Each T-CONT is indexed by an Alloc-ID. T-CONTs represent a logical communication link between the OLT and the ONUs, with every single ONU being able to assigned one or more T-CONTs. There are three different types of T-CONT feasible for dynamic bandwidth allocation, T-CONT types 2, 3 and 4. T-CONT type 2 is for on-off type traffic with well defined rate bound and strict delay requirements provisioned with assured bandwidth. This bandwidth has to be granted to the T-CONTs' traffic, if requested. If not used, bandwidth can be allocated to other T-CONTs, providing that it is available as soon as T-CONT type 2 requires it. T-CONT type 3 is provisioned with assured bandwidth and it can also be granted non-assured bandwidth if the entire assured bandwidth is utilized. It is suitable for variable rate, bursty traffic with requirements for average rate guarantee. T-CONT type 4 has no bandwidth guarantee but it has eligibility in best effort bandwidth sharing.

In XG-PON the DBA follows a strict priority hierarchy among the forms of assigned bandwidth:
1) Fixed bandwidth (highest priority)
2) Assured bandwidth
3) Non-assured bandwidth
4) Best-effort bandwidth (lowest priority)

Firstly, the OLT assigns the upstream bandwidth to the fixed bandwidth of each Alloc-ID. Secondly, the OLT allocates the assured bandwidth of each Alloc-ID as long as the Alloc-ID has enough traffic to consume the assured one. The OLT then satisfies the requirements of non-assured bandwidth to the eligible unsaturated Alloc-IDs until either all of them reach their saturation level, or the surplus bandwidth pool is exhausted. Finally, the OLT allocates the remaining bandwidth to the best-effort bandwidth components.

Because XG-PON adopts this strict hierarchy in bandwidth allocation, the received QoS of a request is determined by the T-CONT type that maps this request. The mapping must consider not only the QoS requirements of the application but also the traffic characteristics. For example, applications such as HDTV streams or video conferencing for business subscribers (who are willing to spend more to acquire a guaranteed QoS) are potentially mapped into T-CONT type 1.

## 3 The 10G-EPON Overview

The effective 1 Gbit/s symmetric data rate supported by the IEEE 802.3-2005 compliant EPON systems has been considered sufficient for a relatively short period of time. However, the increasing demand of raw bandwidth and high capacity, resulted in the development of the 10G-EPON. The 802.3av PON standard [11] was developed to increase the data rate of EPON systems from 1 Gbit/s to 10Gbit/s and being compatible with the 10 Gbit/s Ethernet interface. There are multiple protocols for both 10G-EPON and EPON. As an addition to the Ethernet family of IEEE 802.3, EPON and 10G-EPON layering is very similar to that of Point-to-Point (P2P) Ethernet. The physical layer is connected to the data link layer using the Media-Independent Interface (MII) or the Gigabit Media-Independent Interface (GMII).

### 3.1 Physical Layer

The physical layer specifies the physical characteristics of the optical transceivers. Ethernet has the tradition of adopting mature low-cost designs to promote mass deployment. This philosophy has been the key to the tremendous commercial success of Ethernet. The Physical layer is subdivided into six blocks [12]:
1) MDI specifies the characteristics of the electrical signals which are received from or transmitted to the underlying medium
2) PMD specifies the basic mechanisms for exchange of data streams between the medium and PCS sublayer. The bottom part of PMD contains physical devices, like receiver and transmitter.
3) PMA sublayer specifies functions responsible for transmission, reception, clock recovery, and phase alignment.
4) PCS defines a set of functions which are responsible for converting a data stream received from GMII into codewords, which can then be passed through PMA and PMD and finally transmitted into the medium.
5) GMII specifies a standardized interface between the MAC and PHY layers. This is one of the major interfaces in the 802.3 stack allowing for modular interconnections of various PHY layers to MAC
6) RS maps MAC service primitives into GMII signals, effectively transfers data into PHY and vice versa. In the EPON architecture, RS plays also one more critical role: it is responsible for LLID insertion and filtering all data passing from MAC or PHY.

According to the specifications of 10G-EPON, it offers symmetric 10 Gbit/s at the downlink and the uplink, as well as asymmetric 10 Gbit/s downlink and 1 Gbit/s uplink data rates. Moreover, in 10G-EPON the OLT is equipped with dual rate receivers for 1G or 10G ONUs to be backwards compatible with the existing and widely deployed 1G –EPON. Additionally, the downlink transmission channels are divided for sending data and control information to 1G and 10G ONUs.

Allowing concurrent operation of 1 Gbit/s and 10 Gbit/s EPON systems was a major priority for the 802.3av standard. On the downlink, the 1 Gbit/s and 10 Gbit/s channels are divided based on wavelengths, with the 1 Gbit/s transmission using the 1480 to 1500 nm band and 10 Gbit/s spreads from 1575 to 1580 nm. On the uplink, there is an overlap; 1 Gbit/s uses the 1260 to 1360 nm band, whereas 10 Gbit/s spread from 1260 to 1280 nm.

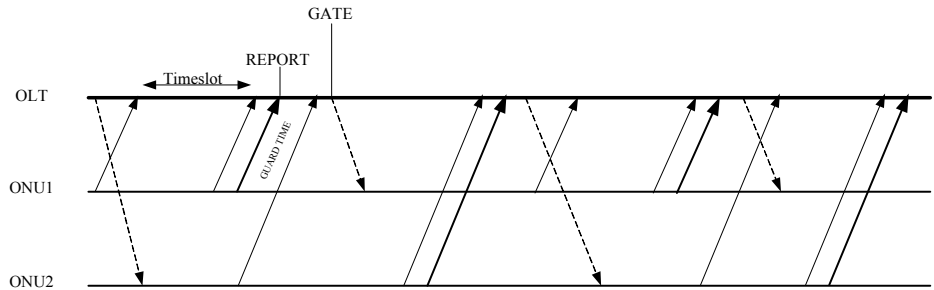

**Fig. 4.** The DBA process in 10G-EPON Systems.

### 3.2 Data Link Layer

The Ethernet layering architecture has differences between P2P and P2MP models. In the data link layer of 10G-EPON, in P2P Ethernet a mandatory Multipoint Media Access Control (MPMC) layer replaces the optional MAC sublayer. The MultiPoint Control Protocol (MPCP) is part of the MPCP layer and is employed to manage access of the 10G-EPON ONUs to the shared PON medium. It is noted that despite the fact that the OLT and ONU stacks are quite similar, MPCP in an OLT has the role of the master, while the MPCP entity in ONU functions as the slave.

MPCP is the protocol adopted to arbitrate the uplink transmission among the ONUs, defined by the IEEE 802.3ah task force. It does not involve a specific DBA scheme, but it allows the implementation of DBA schemes by facilitating information exchange necessary by the OLT to assign bandwidth to each ONU. MPCP controls the access to the P2MP topology architecture via message, status and timer. Its main functions are ONU bandwidth allocation, polling ONU bandwidth requests, reporting congestion to up-level, ONU auto-discovery, registration, and ranging[13,14]. It includes two 64-byte MAC control messages, GATE and REPORT [14].

**Dynamic Bandwidth Allocation in 10G-EPON**

There were no significant changes in the DBA scheme supported by GEPON systems suffered when transiting towards 10 Gbit/s EPONs. Due to the case of co-existence, the emerging 10 Gbit/s EPONs have the DBA operation based on the underlying MPCP sublayer. Thus, the DBA entity is responsible for scheduling two mutually cross dependent EPON systems, which use a common single upstream channel. In downlink, the DBA agent schedules the transmission of the GATE MPCPDUs independently, since WDM multiplexing is used to separate the 1 Gbit/s and 10 Gbit/s paths [13-16]. P2P emulation is achieved by a mechanism, which allows the medium to behave as a collection of P2P links. The emulation depends on tagging Ethernet frames with a unique identifier for each ONU called the Logical Link ID (LLID), which is embedded in the frame preamble. Fig. 4. shows the 10G-EPON DBA process.

With DBA, the OLT assigns the bandwidth to different ONUs based on the data they have to send rather than a static allocation per ONU. The 10G-EPON uses

REPORT messages from the ONUs to inform the OLT of their current bandwidth needs. Their bandwidth needs are reported in terms of the number of characters they have in the different priority queues awaiting upstream transmission. The OLT can also take into account the service level agreements (SLAs) that have been specified for the service flows associated with an ONU. The OLT grants bandwidth to each ONU by sending a GATE message that informs the ONU of the start time and duration of its transmission on the upstream channel.

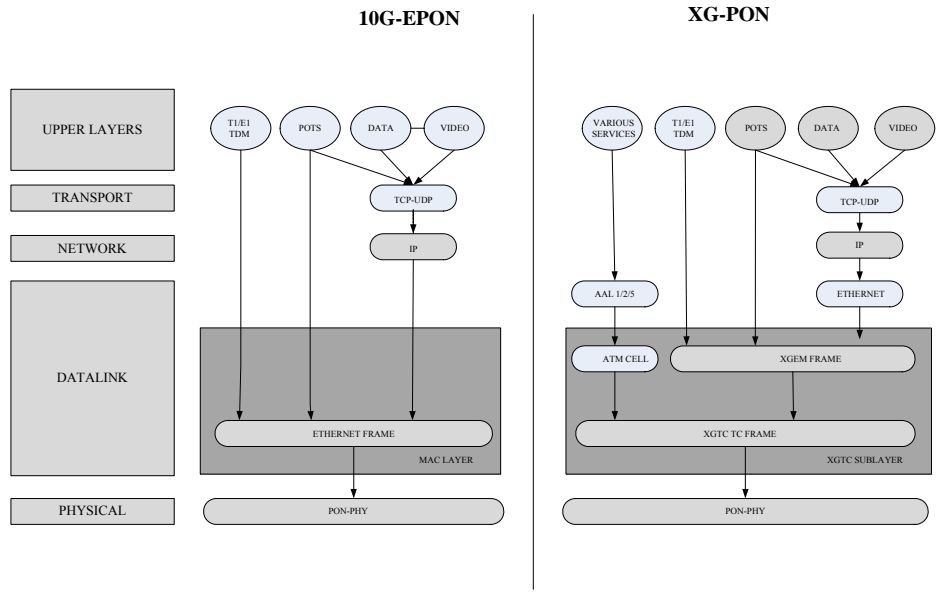

**Fig. 5.** The multilayer comparison between the two standards.

## 4 Comparing the Standards

Both EPON and GPON inherit main standard processes and components from ITU-T G.983 BPON standard. Examples could be found in their general concepts (PON operation, ODN framework, wavelength plan, and application). Nonetheless, significant differences can be found in each modern standard. EPON maintains its Ethernet-based nature incorporating Ethernet protocols. On the other hand, GPON is fundamentally a transport protocol that leverages the techniques of Synchronous Optical NETworking (SONET), Synchronous Digital Hierarchy (SDH) and Generic Framing Protocol (GFP) to transport Ethernet signals. In this Section we try to distinguish the major common features and main differences between the two competent standards. Fig. 5. projects a multilayer comparison approach between the two standards.

### 4.1 Layering and Multiplexing

The 10G-EPON is based upon IEEE 802.3 Ethernet that was modified to support P2MP connectivity. It is able to effectively support all Ethernet features. On the other hand, XG-PON constitutes a more flexible paradigm including generic mechanism for any kind of traffic provided. Its adaptation mechanism is quite powerful and is capable of aggregating multi-source traffic stemming from multiple network structures. In 10G-EPON, Ethernet frames keep their original format and properties, offering a flexible and simple layering model. In essence, Ethernet-based PONs carry out IP-based traffic in an end-to-end fashion. XG-PON encompasses Services are all mapped over Ethernet (directly or via IP). This task takes place in XG-PON systems by engaging two layers of encapsulation. Firstly, TDM and Ethernet frames form XGEM frames and then ATM and XGEM frames are both encapsulated into XGTC frames.

In Ethernet-based PONs the Logical Link ID (LLID) addresses an ONU MAC address with OLT ports. This strategy, also known as P2P emulation concept, bridges user and backbone interfaces. On the other hand, XG-PON utilizes a variable called T-CONT to address each ONU. Both LLID and T-CONTs provide a form of P2P emulation.

### 4.2 Bandwidth Allocation

Reporting the buffer occupancy in the user side is crucial. In the XG-PON, this procedure takes place by inserting it into the fields of the DBRu headers; a single REPORT message is piggybacked in the real data in the 10G-EPON concept. In both cases, the REPORT is mandatory declaring the bandwidth user requests in the uplink direction in terms of Bytes. The transmissions guidelines from the OLT towards the ONUs are provided via the GATE message in Ethernet-based PONs. Usually, a control channel is incorporated to carry out GATE messages. This process is quite different in XG-PON systems. Here, a periodic downstream frame is sent to all ONUs including transmission guidelines for all ONUs in the header. For each ONU the OLT includes the transmission time and the allocated bytes.

In general applying a DBA scheduling algorithm is of paramount importance. However, DBA algorithms are optinal in IEEE 802.3av. The main thinking tank remains the OLT; nevertheless, it's up to the scheduler at the OLT whether / how to construct a transmission schedule. XG-PON employs a very similar scheme, but in that case DBA is part of the standard. Furthermore, the QoS provisioning as a part of the DBA scheme is rigorous and specific.

### 4.3 Bandwidth and Efficiency

One of the most important assets in the 10G-EPON is the symmetric 10 Gbit/s data rates. Both directions support both , as well as asymmetric 10 Gbit/s downstream and 1 Gbit/s upstream, whereas XG-PON provides 10 Gbit/s downstream and 2 Gbit/s upstream. Efficiency though has to be considered in both directions of a PON. By the term efficiency we usually mean throughput efficiency (also called utilization). Throughput is a measure of how much user data (application-level data) the network can carry through in a unit of time. Throughput efficiency is a ratio of maximum throughput to the network bit rate. The need for speed in the downstream direction is significantly important while the upstream efficiency guarantees QoS. Each PON protocol introduced its own overhead in either direction. Overall, PON efficiency is a function of protocol encapsulation and scheduling efficiencies.

Apart from throughput, there are two other major parameters for evaluating a PON performance, latency and fairness. The three criteria are interrelated. Reduced utilization results in increased latency because the available bandwidth to empty an ONT queue is decreased; hence more time is required to empty the queue. Similar impact is noticeable for fairness. Low fairness performance indicates that some ONTs will be served slower than others. Consequently, the latency of the slowly-served ONTs increases. Obviously all of the above depend on the DBA algorithm that is being considered (this is considered as future work of the current paper).

### 4.4 QoS

The QoS provisioning is quite important. Nowadays, the data delivery has been advanced to high-level aggregating multimedia delivery including voice, video, and data capabilities. However, several impairments exist. The QoS concept is realized in the interconnection between the OLT and ONUs. More specifically, the OLT is solely responsible to provide QoS-aware transmission scheduling in both directions without violating the existing SLAs. The PON protocol and architecture provide intercommunication as well as a flow control mechanism that easily facilitates implementation of QoS.

The XG-PON ONU plays a key role in ensuring QoS for all traffic because it is the ingress/egress point for all network traffic. The ONU and therefore all users that are 'hooked' behind it, can experience all types of modern applications. In Ethernet-based PONs this is accomplished by applying strong IP-based QoS support. The ONU can also perform service classification based on the physical port and map it to 802.1p p-bits. For example, traffic flows from voice ports can be classified as highest priority. As part of this service differentiation, the ONU associates different traffic flows with a specific XGEM Port ID. These are virtual port identifiers that have significance for a given XG-PON. In 10G-EPON the QoS provisioning dramatically depends on the allocation mechanism, i.e., the MPCP. There is a large variety of DBA algorithms proposed to use the multipoint control protocol defined in order to arbitrate the transmission of different users. Several DBA algorithms support QoS support in native.

## 5 Conclusions and Open Issues

The scope of the current paper was to study and compare the two major standards ITU-T XG-PON and IEEE 10GE-EPON for next generation passive optical networks and technologies . The 10G-EPON design aims at exploiting the widespread and mature Ethernet technology for reducing component development effort, design cycles and overall cost. On the other hand, XG-PONs aims at higher line rates accepting higher receiver circuit costs while targeting a set of mechanisms for flexible traffic multiplexing, detailed traffic management specifications and QoS guarantees with better control of network resource allocation as well as operation and maintenance.

The basic differences of these two technologies were already known from the wide deployment of their predecessors, EPON and GPON. The fundamental difference between 10G-EPON and XG-PON is that XG-PON is a transport technology for Ethernet as well as TDM and ATM. Moreover, GPON utilizes an out-of-band bandwidth allocation map with the concept of traffic containers as the upstream-granted entity. The services are encapsulated into frames in their native format by a process called XGEM while EPON uses no encapsulation (thus, Ethernet traffic is transported natively and all Ethernet features are fully supported). The key aspect of XG-PON's low-latency capability is that all upstream TDMA bursts from all ONUs can occur within 125 μsec. Each downstream frame includes an efficient bandwidth allocation on the BWmap field of the header of XGTC frame, which specifies a bandwidth allocation to a particular Alloc-ID broadcasted to all ONUs and can support a fine granularity of bandwidth allocation. This out-of-band mechanism enables the GPON DBA to support very small grant cycles without compromising bandwidth utilization. XGEM also supports fragmented payloads, which are not allowed in 10G-EPON standard. Thus, a low Class of Service (CoS) T-CONT can stop its upstream burst in the middle of a payload, allow a higher CoS T-CONT its access, and then resume its transmission when told to by the DBA mechanism. Thus, in a highly utilized PON, large bursts of low priority, best-effort data will have minimal affect on high priority, delay-sensitive traffic like voice and TDM. On the upstream direction the two standards work pretty much with the same philosophy. The 10G-EPON uses the MPCP REPORT message and XG-PON the BWmap field in every upstream burst. In that way every ONU lets the OLT know the next desired bandwidth allocation. Of course, the overall merit of the two competitors cannot be judged on technology terms alone. Performance, efficiency and cost have a crucial bearing on deciding which one is the next best thing.

In terms of deployment, although EPON and GPON are widely deployed worldwide, (GPON mostly in North America while EPON in Asia and Europe), there are very few deployments of next generation 10G-EPON and XG-PON. The major reason for this is the relatively high deployment cost for 10G PON which is still in an early state.

Knowing the many benefits of high capacity networks, such as higher bandwidth, enhanced QoS, more bandwidth for each subscriber, lower cost, etc, IEEE points out that will investigate requirements for the next generation of EPON through the activities of the IEEE Standards Association (IEEE-SA). The initiative will attempt to measure the need to support data rates beyond the current top speed of 10 Gbits [17]. Despite deployments of the technology haven't yet taken off, IEEE has decided to test the waters regarding the next set of EPON specifications, as "equipment vendors and network operators, particularly in Asia and North and South America, are interested in exploring the technologies available for the next generation of EPON," according to an IEEE press release [17].

In conclusion it can be seen that both the IEEE and the ITU-T recognized the need to evolve gigabit PONs to 10Gbit/s capable solutions for transporting Ethernet and IP traffic. Due to the need for more bandwidth, they developed 10G-EPON and XG-OPN, respectively. Although, these two standards are very different in execution, concerning Ethernet transport, management, bandwidth allocation etc, they are equally capable of providing valuable QoS capabilities required for triple play services, each one in a different way.